\documentclass[12pt,preprint]{aastex}

\gdef\etal{{et al.}}

\gdef\ltsima{$\scriptscriptstyle \; \buildrel < \over \sim \;$}
\gdef\simlt{\lower.3ex\hbox{\ltsima}}

\gdef\gtsima{$\scriptscriptstyle \; \buildrel > \over \sim \;$}
\gdef\simgt{\lower.3ex\hbox{\gtsima}}

\shorttitle{Background Subtraction in 2D Spectroscopy}
\shortauthors{Kelson}
\slugcomment{Accepted for Publication in PASP}

\begin{document}

\title{Optimal Techniques in Two-dimensional Spectroscopy:
Background Subtraction for the 21st Century}

\author{Daniel D. Kelson}
\affil{Carnegie Observatories, 813 Santa Barbara Street, Pasadena, CA 91101}
\email{kelson@ociw.edu}

\begin{abstract}
In two-dimensional spectrographs, the optical distortions in the spatial
and dispersion directions produce variations in the sub-pixel sampling
of the background spectrum. Using knowledge of the camera distortions
and the curvature of the spectral features, one can recover information
regarding the background spectrum on wavelength scales much smaller than
a pixel. As a result, one can propagate this better-sampled background
spectrum through inverses of the distortion and rectification
transformations, and accurately model the background spectrum in
two-dimensional spectra for which the distortions have not been removed
(i.e. the data have not been rebinned/rectified). The procedure, as
outlined in this paper, is extremely insensitive to cosmic rays, hot
pixels, etc. Because of this insensitivity to discrepant pixels, sky
modeling and subtraction need not be performed as one of the later steps
in a reduction pipeline. Sky-subtraction can now be performed as one of
the earliest tasks, perhaps just after dividing by a flat-field. Because
subtraction of the background can be performed without having to
``clean'' cosmic rays, such bad pixel values can be trivially identified
after removal of the two-dimensional sky background.
\end{abstract}


\section{Introduction}

For more than 100 years, optical astronomers have employed long-slit
spectrographs to study the internal physics of heavenly objects. Such data
contain the target's spectrum dispersed at every location along a single
position angle on the sky (unless the target is an unresolved source).
Unfortunately, at every location along the slit, one also collects photons
from the night-sky emission. This background must be removed from the data
in order to reveal the spectrum of the intended target. With the expansion
of spectroscopy to wide fields-of-view, one can employ multi-slit aperture
plates to collect spectra for many objects simultaneously, or observe with
a very long slit. Over the past several years, important advances have been
made in the art of background subtraction, though largely in the area of
small-aperture spectroscopy, either with fibers or ``micro-slits''
\citep[e.g.,][]{svdfit,nod,viton}. However, or typical long- and multi-slit
spectroscopy, the most common reduction procedures do not make optimal use
of the data, and the resultant spectra with which one must perform one's
science suffer in quality compared to what is achievable with more modern
techniques and modest computing power.

This paper outlines a new technique which makes optimal use of the data to
accurately perform the subtraction of the unwanted background spectrum from
two-dimensional spectra {\it before\/} any rebinning of the data is
performed. In the procedure one makes full use of the spectrograph
distortions to improve the sampling of the sky background spectrum. The
quality of the background subtraction is completely insensitive to the
magnitude of the distortions imposed by the spectrograph's camera, to the
severity of the curvature of the spectral lines that is caused by the
dispersive element, or even to any tilting of individual slitlets in an
aperture mask. Furthermore, by explicitly employing maps of the
y-distortion and line curvature in a two-dimensional spectrum, the model
two-dimensional background spectrum not only follows the same line
curvature and y-distortion as the data, but the spectral features are
sampled (pixelated) in exactly the same way as the features are sampled in
the raw observations. As a result, when one subtracts the model from the
data, there are absolutely no sharp residuals at the edges of night-sky
emission lines. With more traditional methods, two-dimensional spectra must
be rectified before one can perform the subtraction of the background; such
rectification procedures introduce artifacts into the data, particularly
when the sampling is poor. These artifacts can manifest themselves as sharp
residuals at the edges of features with strong gradients, e.g., the night
sky lines. Furthermore, in traditional methods, observers are required to
identify (and remove) cosmic rays and other bad pixels before rebinning the
data. With the method discussed in this paper, the sky subtraction is
performed before the data have been rebinned, and, as a result, cosmic rays
can be ``cleaned'' after the task of removing the sky background.

In the following sections, the method for fitting the two-dimensional
night-sky background is described. Subsequently, this powerful new
technique is applied to data riddled with cosmic rays and bright night-sky
emission lines. The examples include data from three spectrographs, in
which the data span a range of sampling from marginal (LRIS),  to slightly
under-sampled (NIRSPEC), to grossly under-sampled (MIKE). And finally the
the advantages of this method over traditional methods will be summarized.

\section{The Basic Procedure}
\label{sec:proc}

Two-dimensional spectra, when ultimately imaged onto a detector, suffer
from two problems that must be dealt with before or during the process
of background subtraction: (1) the fact that the two-dimensional spectra
are not aligned exactly along the rows (or columns) of the detector and
are often curved with respect to the natural coordinate system of the
detector (the y-distortion); and (2) the general tendency for dispersers
to impose a wavelength-dependent line curvature onto the two-dimensional
spectra (which may have already been tilted or curved if the slit has
been so cut into the aperture plate). Furthermore, the coarse pixel
sizes of modern detectors impose a third problem which normally limits
the accuracy with which one could previously deal with the first two.

In order to discuss these issues and their resolution, we first define the
image of two-dimensional spectroscopic data as $P(x,y)$, where $(x,y)$ are
pixel coordinates in the system of the original image (e.g., a CCD frame).
Because of distortions imposed by the optics, the spatial coordinate on the
sky, $y_t$, for a given pixel $(x,y)$ is a non-linear function, $y_t =
Y(x,y)$. Furthermore, the wavelength of light, $\lambda$, incident onto a
pixel $(x,y)$ is also a non-linear function of image position. For the
purposes of modeling the two-dimensional background spectrum, we are less
concerned with the actual wavelength, $\lambda$, of incident light than we
are with the fact that there exists a coordinate system, $(x_r,y_t)$, in
which $x_r$ is a wavelength-dependent coordinate that is orthogonal to the
spatial coordinate $y_t$. The transformation to this system is $x_r =
X(x,y_t)$ such that the wavelength of light incident on a pixel can be
written $\lambda = L(x_r)$, where $x_r = X(x,Y(x,y))$. Thus there exists a
convenient coordinate system $(x_r,y_t)$ for which $\partial L /\partial
y_t=0$. The transformations $Y(x,y)$ and $X(x,y_t)$ can be measured with
great precision from comparison lamp spectra or from the night-sky features
themselves \citep[see, e.g.,][for a description of a robust algorithm using
FFTs and cross-correlations in order to make full use of the available data
to map these distortions]{expector}\footnote{Note that that exact knowledge
of the distortions does not free the observer from artifacts imposed on
one's data by the process of interpolation.}.

With knowledge of $Y(x,y)$ and $X(x,y_t)$, the traditional method of
sky-subtraction required one to interpolate $P(x,y)$ onto a regular grid
in $(x_r,y_t)$ before fitting the two-dimensional sky spectrum at every
individual interval of $\lambda$ in the rebinned image. Unfortunately,
the process of rebinning the data (1) introduces correlated noise; (2)
smears cosmic rays and other bad pixels which might not have been
flagged/cleaned beforehand; and (3) produces artifacts at the edges of
sharp features. This last problem sometimes leads users to invoke high
spatial orders in the fitting of the sky background in order to subtract
such artifacts at the edges of sky lines. Furthermore, the rebinning of
the data forces every sky spectrum used in the fit to have a common
pixelation. This process limits one's ability to accurately model the
two-dimensional sky spectrum where gradients (e.g., $\partial P/\partial
x_r$) can be large.

Instead of rebinning the data before performing the modeling of the
two-dimensional sky spectrum, we propose that one should perform a
least-squares fit to the sky spectrum using the original data-frame, in
which the distortions and line curvature have not been removed. Using
$y_t = Y(x,y)$ and $x_r = X(x,y_t)$ one should model the background
spectrum in $P(x,y)$ as a function of the rectified coordinates
$(x_r,y_t)$.

\begin{figure}
\plotone{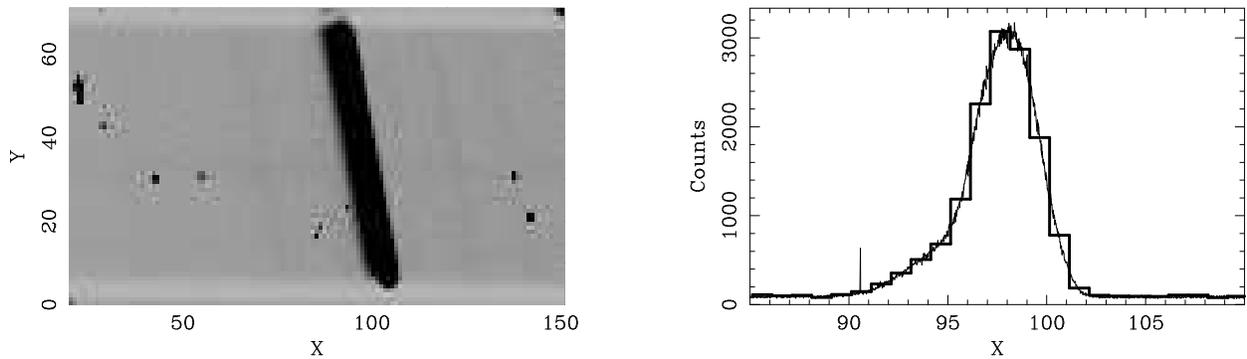}
\caption{(left) A subsection of an LRIS two-dimensional spectrum surrounding
the 5577\AA\ night-sky emission line. (right) The thick line shows the
intensity of the 5577\AA\ line from a single CCD row. The thin line
shows the value of every pixel in the left-hand panel, plotted as a
function of its rectified position, $x_r=X(x,y_t)$, along the
wavelength-dependent coordinate in the rectified coordinate system. Note
that the shape of the line is actually quite well sampled as a result of
the tilt of the spectral line, due to both the tilting of the slit on
the aperture plate and the line curvature imposed by the instrument's
grating. If the data were to be rebinned, this over-sampling would be
lost. Note the spike at $x\approx 91$, at which a cosmic ray is clearly
visible above the rest of the data (see Figure
\ref{fig:section}).
\label{fig:5577}
}
\end{figure}

Each pixel in the original image represents an integral of the flux
within a box the size of 1 pixel, but in the analysis each observed
pixel's location is assumed to be $(x_r,y_t)$, rather than $(x,y)$. In
this way, each pixel samples the sky spectrum at a {\it known\/}
sub-pixel position. Figure \ref{fig:5577}, in which a small section of
an LRIS \citep{okelris} spectrum is shown, demonstrates the utility of
this change in coordinate systems. The left-hand panel shows a region
around 5577\AA\ in a short two-dimensional spectrum obtained using the
600 g/mm grating ($\sim 1.28$\AA/pixel). The resolution is $\sim 3.5$
pixels (FWHM). Note in the image how pixelated the edges of the sky line
are. Rebinning such data would lead to spatially periodic artifacts
along the edge of the line. In the right-hand panel, the thick line
shows the spectrum from one row, indicating how pixelated and poorly
sampled the gradients in the line profile are in a given CCD row. The
thin line shows the pixel values in the image $P(x,y)$ as a function of
$x_r=X(x,y_t)$, revealing how well the CCD frame samples the sky
spectrum. Such over-sampling is lost when one rebins the data.

\begin{figure}
\plotone{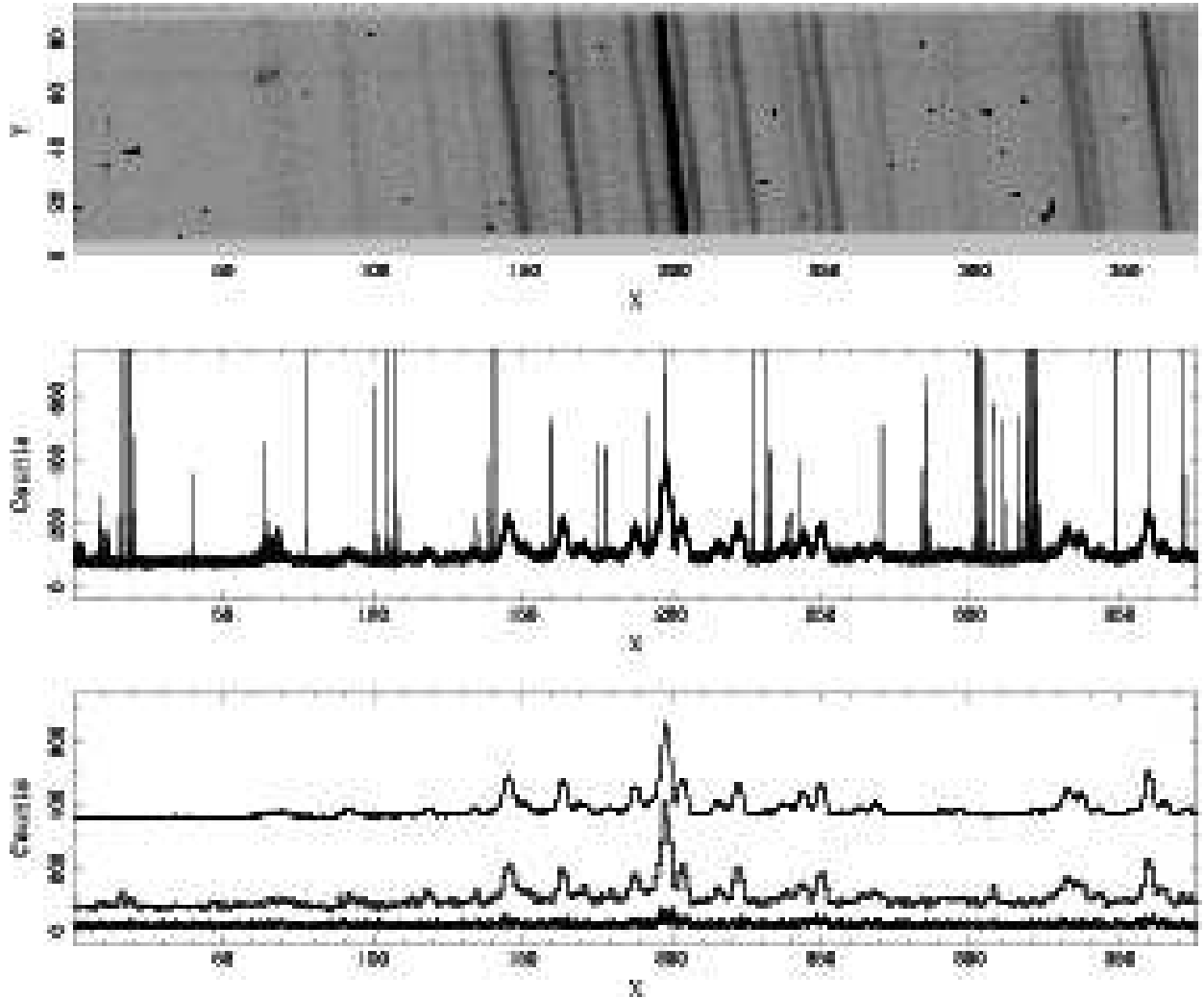}
\caption{(top) A subsection of an LRIS two-dimensional spectrum surrounding
the 6300\AA\ night-sky emission line. (middle) The raw spectrum as
sampled in every row of the image section, where the pixel values are
plotted as a function of $x_r$ (the rectified wavelength-dependent
coordinate). Note that, as in Figure \ref{fig:5577} the night-sky
emission lines are well-sampled as a result of the distortions and line
curvature. Also note that many spikes in the data, representing cosmic
rays and other bad pixels, are clearly visible. (bottom) The thick line
shows the spectrum from a single row, indicating the coarse sampling by
the LRIS pixels. Artificially shifted higher is the 30\%-smoothed
version of the spectrum shown in the middle figure. The line hovering
near zero in the bottom panel shows the $4\times \sigma$ scatter as a
function of $x_r$. Using the percentile-smoothing of the data along the
$x_r$ direction together with a robust $\sigma$-clipping algorithm
allows one to reject cosmic rays and bad pixels from the fitting of the
B-spline.
\label{fig:section}
}
\end{figure}

In Figure \ref{fig:section}, a larger section of the same LRIS frame is
shown in the top panel. The middle panel shows $P(x,y)$ plotted as a
function of $X(x,y_t)$. Note that many spikes in the data are visible.
These discrepant points correspond to cosmic rays and other bad pixels.
Fortunately, the sky spectrum is actually quite over-sampled in the
rectified coordinate system, with many redundant measurements of the sky
at nearly the same sub-pixel location.

Because of the redundancy in the sampling of the background spectrum, it
becomes straightforward to identify the cosmic rays and bad pixels, even
when they appear where gradients in the background spectrum are large.
In the bottom panel, the thick line shows the spectrum sampled by a
single row. The smooth line, shifted towards positive values, is a
smoothed version of the data in the middle panel. The smoothing that was
adopted was a running 30th-percentile within a window 30 data elements
wide in $x_r$. In the bottom panel, we also show the running $4\sigma$
scatter (robustly determined) within the same sliding window. By
comparing $P(x,y)$ with the smoothed spectrum in the middle panel, one
can employ a simple $\sigma$-clipping algorithm to reject the cosmic
rays and bad pixels, even where there are strong gradients in $P(x,y)$.
Normally, the rejection/identification of cosmic rays or bad pixels
involves the comparison of a pixel in an image with its nearest
neighbors. However, because of the poor sampling of strong gradients in
the background spectrum, one should compare a given pixel to those
nearest only in $x_r=X(x,Y(x,y))$, i.e. sampled at the same sub-pixel
interval. Thus, the smoothing is done on a copy of the array $P(x,y)$
that has been sorted in order of increasing $x_r$. Because the line
curvature (plus the additional tilt of the slitlet) has improved the
sampling of the night-sky spectrum, cosmic rays that fall where there
are strong gradients in the sky spectrum can still be robustly
identified, as other CCD rows have sampled the sky spectrum with similar
sub-pixel sampling as at the location of a given cosmic ray.

\begin{figure}
\plotone{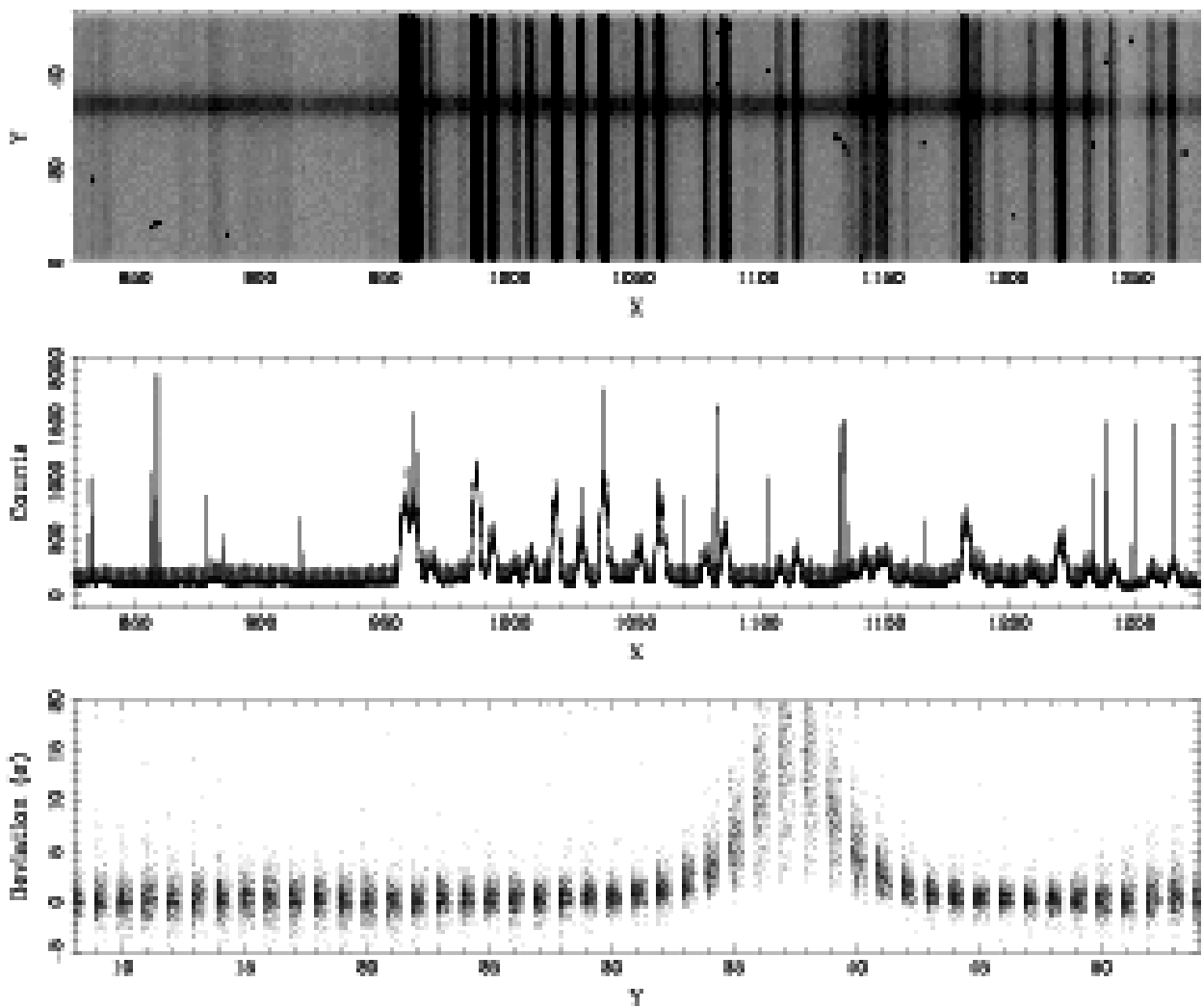}
\caption{(top) A subsection of an LRIS two-dimensional spectrum in the
red, with a bright object in the slit. (middle) The spectrum as sampled
in every row of the image section shown, where the pixel values are
plotted as a function of $x_r$ (the wavelength-dependent coordinate in
the rectified coordinate system). Note that while many cosmic rays are
clearly visible, there also appear to be spikes peaking above the data
at regular intervals in $x_r$. These discrepant pixels have counts which
are dominated by flux from the object. (bottom) The deviations from the
percentile-smoothed data plotted as a function of spatial position along
the slit. Note that the same $\sigma$-clipping algorithm that rejects
cosmic rays also rejects pixels contaminated by objects.
\label{fig:object}
}
\end{figure}

While cosmic rays and other bad pixel values tend to be isolated sharp
features, objects are generally extended in $(x_r,y_t)$. The top panel
of Figure \ref{fig:object} shows a subsection of data for a slitlet
containing a bright object. The middle panel shows every pixel $P(x,y)$
plotted, again, as a function of $X(x,y_t)$. Most of the pixels in the
data only contain flux from the night-sky background, and these data
points follow a locus in the figure with very small scatter. As in
Figure \ref{fig:section}, the cosmic rays are clearly visible above the
well-defined background spectrum. Also visible, however, is a collection
of points in the data which peak above the sky spectrum at regular
intervals of $X(x,y_t)$. If we take the residuals of $P(x,y)$ from the
robustly-smoothed version of the sky spectrum in the middle panel, and
plot the statistical significance of those residuals against the spatial
coordinate $y_t=Y(x,y)$, as in the bottom panel of the figure, the
object pixels are clearly visible as a significant positive deviation
from zero. By employing the $\sigma$-clipping described above for
flagging cosmic rays, one also singles out those pixels that are
significantly contaminated by flux from the object. In general a choice
of clipping at 3 or $4\sigma$ is very effective at rejecting cosmic rays
from the fit to the sky background, and it also rejects objects which
could seriously affect the fit to the sky. Faint objects, which do not
peak above the adopted threshold tend not to affect the modeling of the
sky and most such objects tend to cover too small a spatial area to
adversely affect the fit anyway. While a clipping method works
sufficiently well for most applications, one can straightforwardly
implement an input set of sky apertures, outside of which pixels are
simply ignored. Sky apertures may be specifically useful in very short
slitlets, where the the sky covers $\simlt 10\%$ of the spatial extent
of the slit.

Once the discrepant pixels are rejected, a bivariate B-spline
\citep[e.g.,][]{splinebook,dierckx} is constructed as an approximation to the
remaining remaining data points as a function of their positions.
However, the pixel values are not considered to represent a bivariate
function of $(x,y)$ (the original CCD coordinates) but the data are
treated as a bivariate function of $(x_r,y_t) = (X(x,Y(x,y)),Y(x,y))$.
In the author's implementation of the method, the DIERCKX surface- and
curve-fitting library available from NETLIB was used. This library
allows one to weight each pixel during the fit for the B-spline
coefficients, and these weights were set equal to the inverse of the
expected noise.

When fitting for the B-spline representation of a bivariate dataset, the
smoothness of the model is set by the density of knots in the two
cardinal directions. In the simplest construction of the two-dimensional
sky background, the knot locations, $t_x$ and $t_y$, are chosen with a
high density in $x_r$ such that $t_x = \{{\rm min}(x_r),{\rm
min}(x_r)+1,\ldots,{\rm max}(x_r)\}$ with intervals of $\Delta x_r=1$.
The knots in $y_t$ are chosen to be $t_y = \{{\rm min}(y_t),{\rm
max}(y_t)\}$, where the minima and maxima in $x_r$ and $y_t$ are derived
for the slit being analyzed. In this way, the B-spline is non-parametric
function in $x_r$ and a polynomial of order $k_y$ in $y_t$. The choice
of $k_y\in \{1,3,5,\ldots\}$ sets the order of the spatial variation in
the sky spectrum at a fixed $x_r$ (e.g., fixed $\lambda$).

With the default choice of knots in $x_r$ described in the previous
paragraph, the fit to the data very nearly approximates an interpolating
spline along the wavelength-dependent coordinate. However, the B-spline
was generated using data with finer sampling than that available in a
single CCD row. As a result, the spline is a smooth representation of
the sky spectrum at those locations $(x_r,y_t)$ in the original data
frame. One can reduce the number of knots in $t_x$ to impose greater
smoothness on the model sky spectrum in those ranges of $x_r$ with
little structure, while leaving a higher density of knots near bright
night-sky emission lines in order to better match the gradients there.
While such optimization of the knots in $t_x$ can improve the background
modeling with particularly noisy data, the selection of $t_x$ described
above is generally sufficient. One can also insert more knots in $t_y$
to model more complicated spatial variation in the sky (at fixed
$\lambda$). With clever placement of knots in the spatial direction, the
model can better map high-frequency spatial variations such as in data
containing residual fringes in the very red portions of a spectrum. If
the spatial variation in the sky spectrum is expected to be negligible,
such as in very short slits, a one-dimensional B-spline can be fit to
the values of $P(x,y)$ as a univariate function of $x_r$.\footnote{One
optimization for $t_x$ in the bivariate method includes first finding
the optimal univariate B-spline representation of the sky spectrum, with
the resulting optimal knot locations subsequently used in the bivariate
fit to the data.}

Regardless of the complexity of the knot placement, the B-spline
representation of the data is well-behaved because every pixel location
in the original image, $(x,y)$ has a correspondingly unique, rectified
coordinate $(x_r,y_t)$. As a result the bivariate fit is well-defined
and computationally inexpensive to construct. Once the B-spline
representation has been computed, it can be quickly evaluated at every
$(x_r,y_t)$ location in the spectrum being analyzed. In this way, the
two-dimensional background spectrum is only evaluated at the locations
where the original pixels exist. Therefore no interpolation at sub-pixel
locations (in $x,y$) occurs in the original data frame. As a result,
there are no artifacts at strong gradients.

Some examples of the application of this method of optimal background
subtraction are shown in the next section. The examples shown below used
the simplest knot selection in the above discussion. No optimization of
knot location was implemented. Finally, the examples were generated
using $k_x=3$ and $k_y=1$.

\section{Examples}

\subsection{LRIS}

\begin{figure}
\plotone{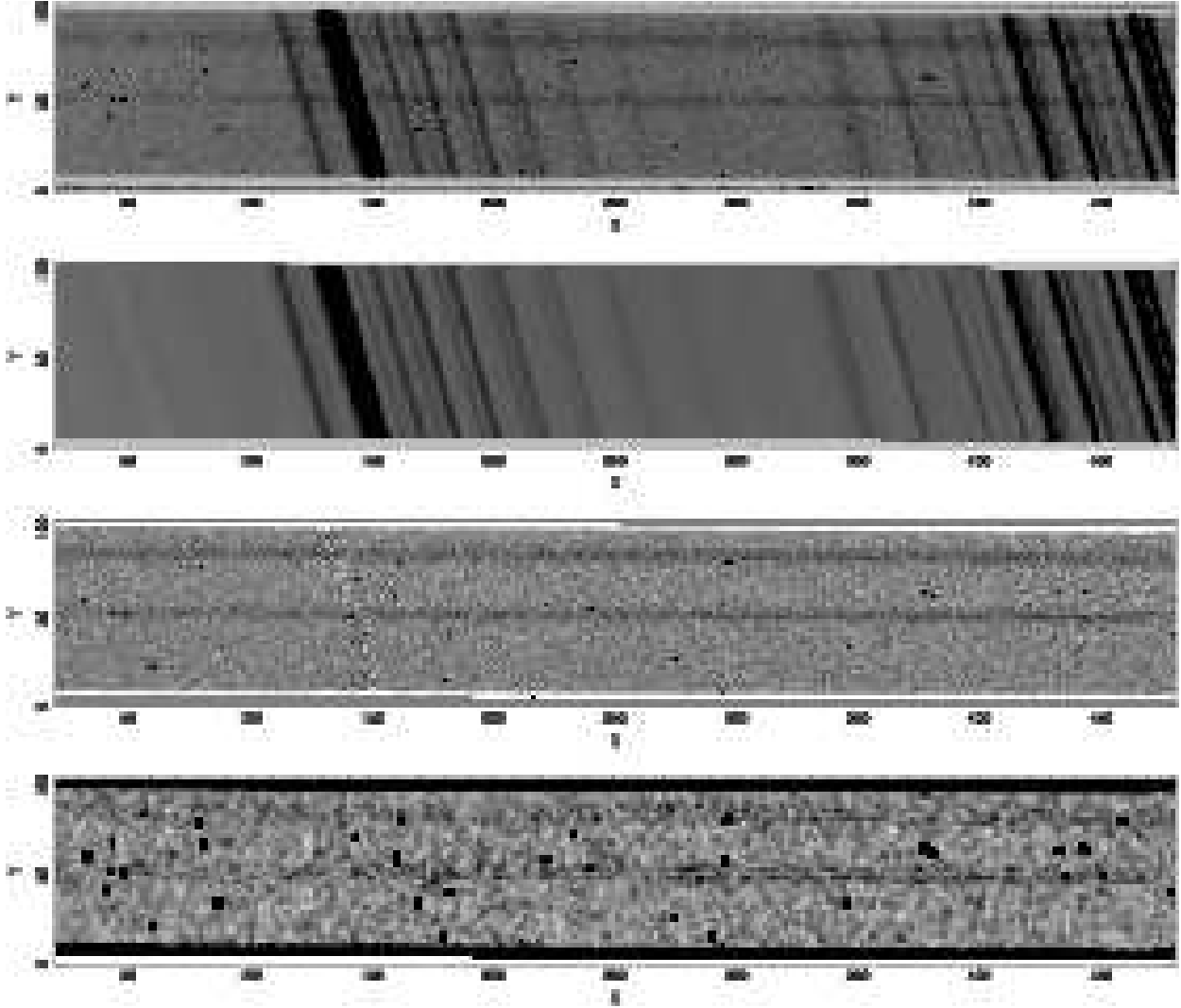}
\caption{(top) A subsection of an LRIS two-dimensional spectrum covering
from Na I to 6300\AA. (top-middle) The two-dimensional fit to those
pixels remaining after the $\sigma$-clipping. (bottom-middle) The
background-subtracted spectrum. (bottom) An $rms$-smoothed image of
the background-subtracted spectrum, divided by the expected noise
(photon and read noise). Other than at the locations of the cosmic rays,
the noise in the final sky-subtracted two-dimensional spectrum has no
additional artifacts or unexpected features, such as would have arisen
at the edges of rebinned night-sky emission lines.
\label{fig:subtract}
}
\end{figure}

The top panel of Figure \ref{fig:subtract} shows a section of
two-dimensional spectrum from one strongly tilted slitlet. The
wavelength range covers from blueward of Na I to approximately 6300\AA.
The second panel shows the model sky spectrum, derived using the new
method. The third panel shows the difference between the two (i.e., the
background-subtracted spectrum). Note how well the night-sky emission
lines are subtracted, with no residual systematic structure. The bottom
panel shows an $rms$-smoothed version of the background-subtracted
image, normalized by the expected noise. This representation of the data
illustrates that there is no additional noise at the locations of the
sharp sky lines.

\begin{figure}
\plotone{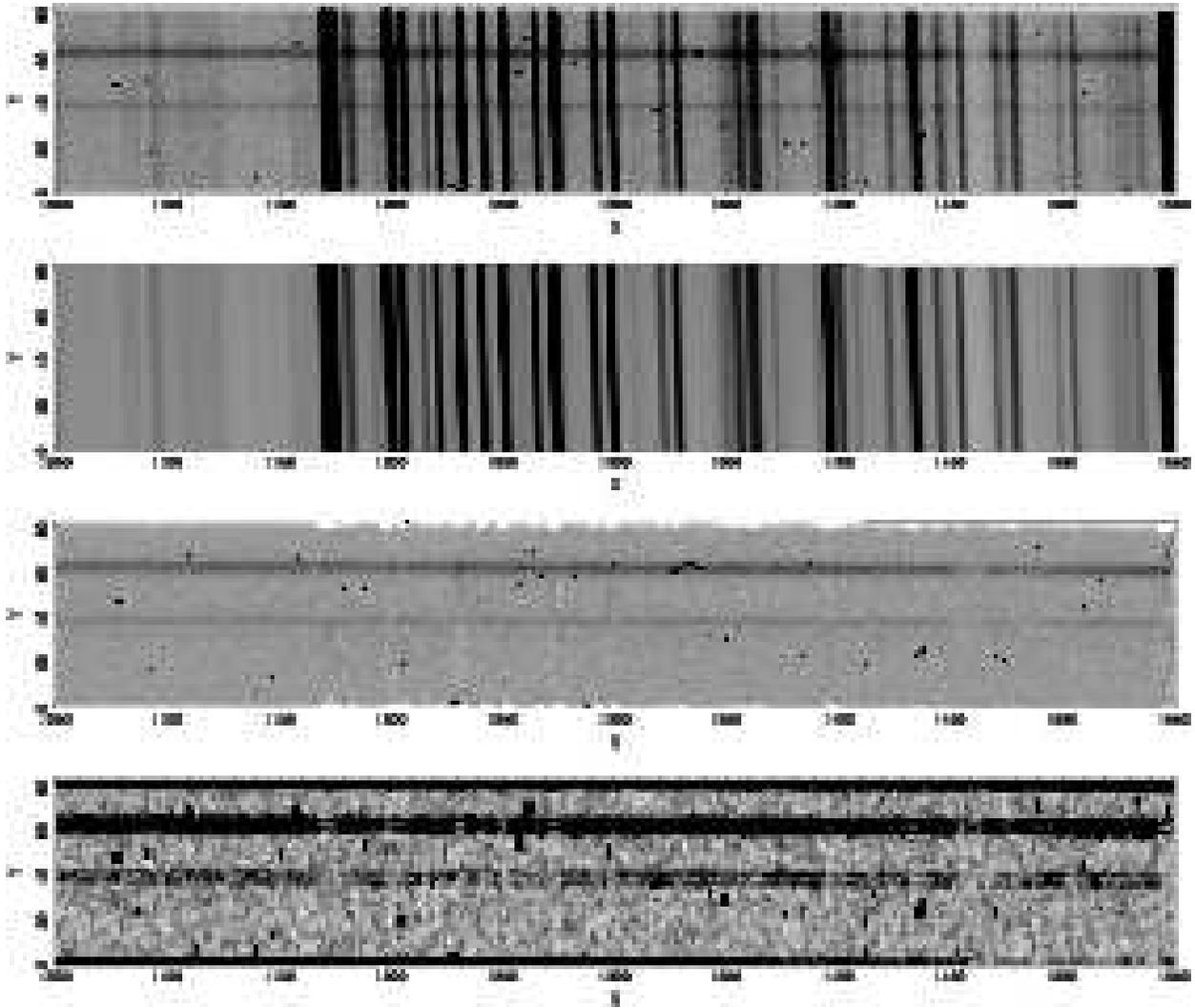}
\caption{Same as in Figure \ref{fig:subtract} but for a different
slitlet, between 7000\AA\ and 7700\AA.
\label{fig:subtract2}
}
\end{figure}

Figure \ref{fig:subtract2} follows a similar format as Figure
\ref{fig:subtract} but for a different slitlet, and for a wavelength
range 7000\AA\ $\simlt \lambda \simlt$ 7700\AA. As in the previous
figure, the background subtraction is free of any residual systematic
structure, and the noise in the background-subtracted image is as
expected.

\begin{figure}
\plotone{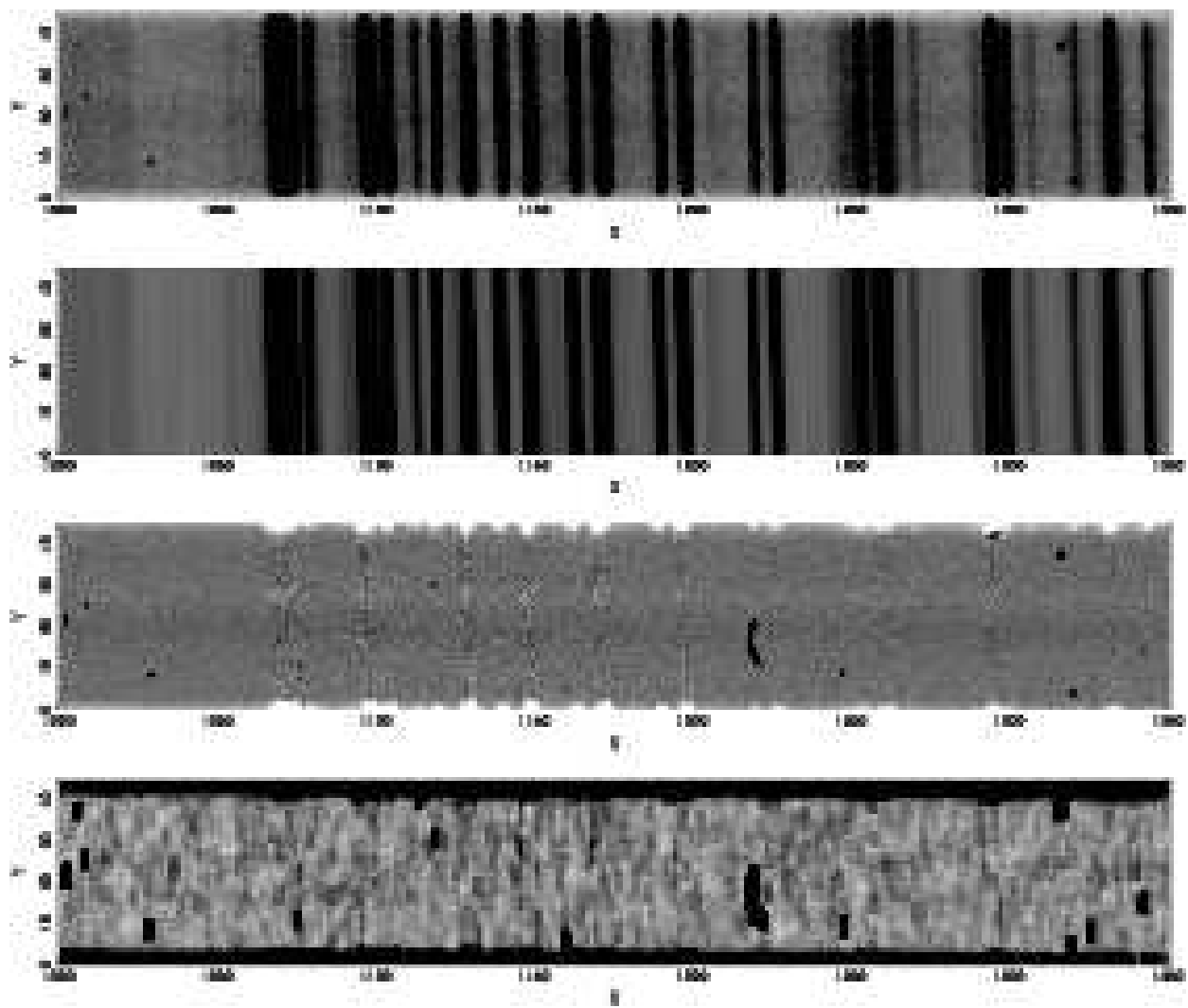}
\caption{Same as in Figure \ref{fig:subtract} but for a short slitlet
in which a serious cosmic ray sits along the top of a sky line. The
robust $\sigma$-clipping algorithm easily handles such nasty occurrences.
\label{fig:subtract3}
}
\end{figure}

Figure \ref{fig:subtract3} shows a section of data with a wavelength
range 7200\AA\ $\simlt \lambda \simlt$ 7600\AA. This section has a very
strong cosmic ray at $(x,y)\approx (1220,15)$ that sits along a large
portion of a sky line. Note how clean the sky-subtracted image is,
again, with no systematic residuals at the locations of bright lines.

\begin{figure}
\plotone{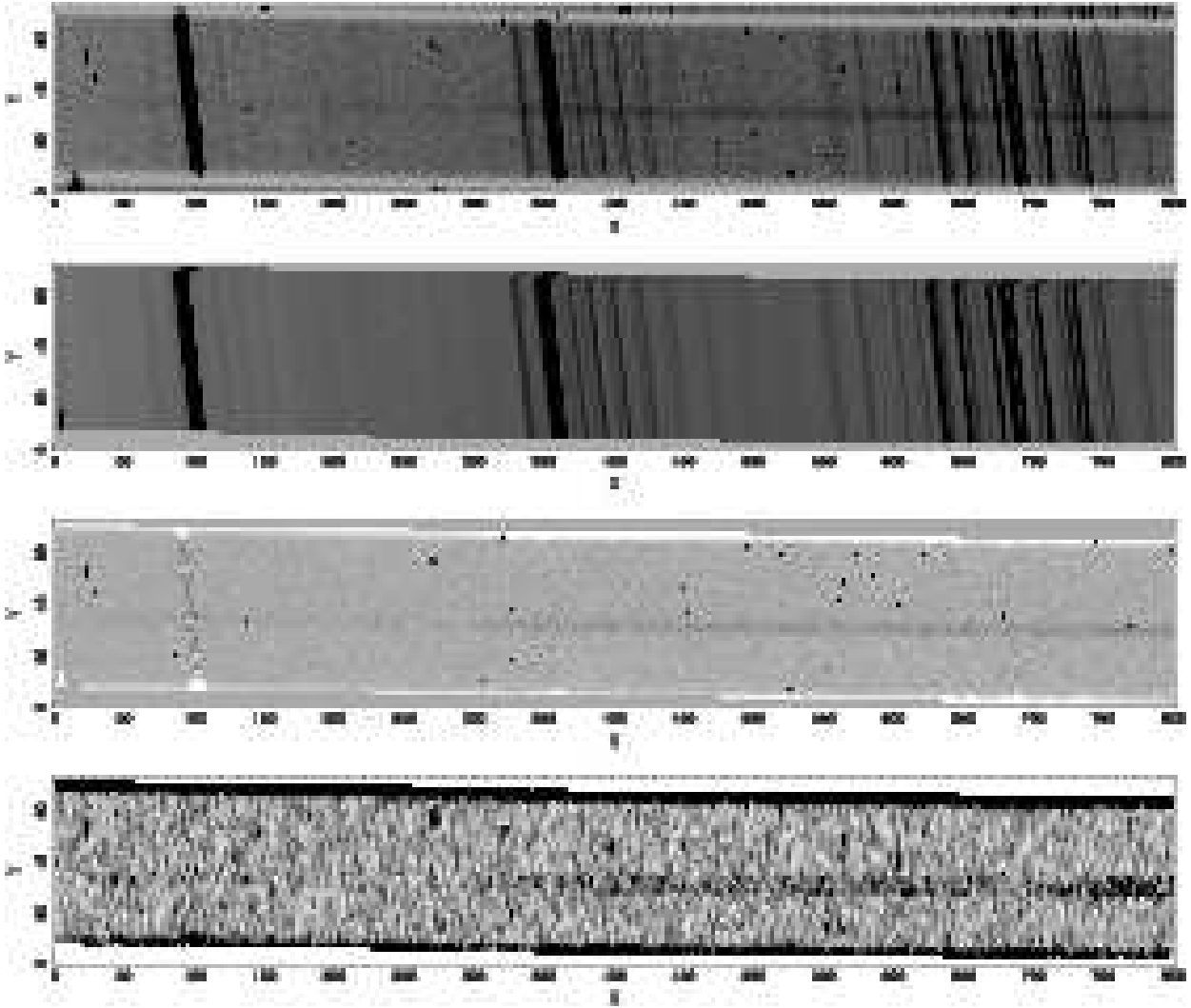}
\caption{Same as in Figure \ref{fig:subtract} but for a tilted slit
covering from 5500\AA\ to 6500\AA. Also note the strong gradient in
the $y$-distortion with wavelength ($\partial Y/\partial x$). Note how
cleanly 5577\AA\ is subtracted, leaving only the expected level of
Poisson noise.
\label{fig:subtract4}
}
\end{figure}

Figure \ref{fig:subtract4} shows a section of data with a wavelength
range 5500\AA\ $\simlt \lambda \simlt$ 6500\AA. These data were taken
from a slitlet at the edge of the LRIS field, at which the
$y$-distortion, $Y(x,y)$, has a very strong derivative with respect to
$x$. Note how cleanly the bright sky lines 5577\AA, 5890\AA, 5896\AA,
6300\AA, etc., are modeled and subtracted. As in the previous examples,
there are no systematic residuals at the locations of bright night-sky
features.

\subsection{NIRSPEC}

\begin{figure}
\epsscale{0.6}
\plotone{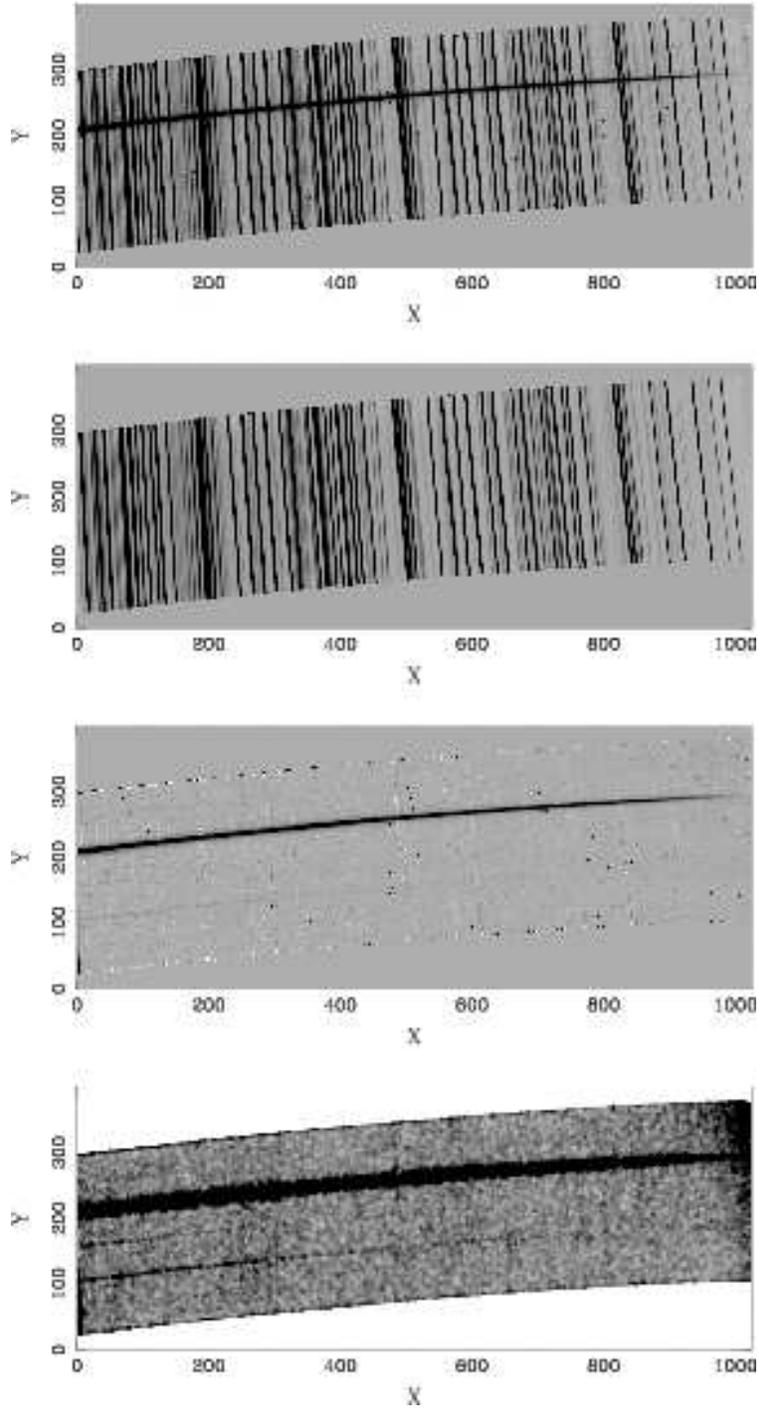}
\epsscale{1.0}
\caption{A single $H$-band long-slit spectrum obtained with NIRSPEC. The
distortions are quite large, and the sampling is poor. Note how cleanly
the night sky emission lines are subtracted, leaving only the expected
level of Poisson noise.
\label{fig:nirspec1}
}
\end{figure}

\begin{figure}
\plotone{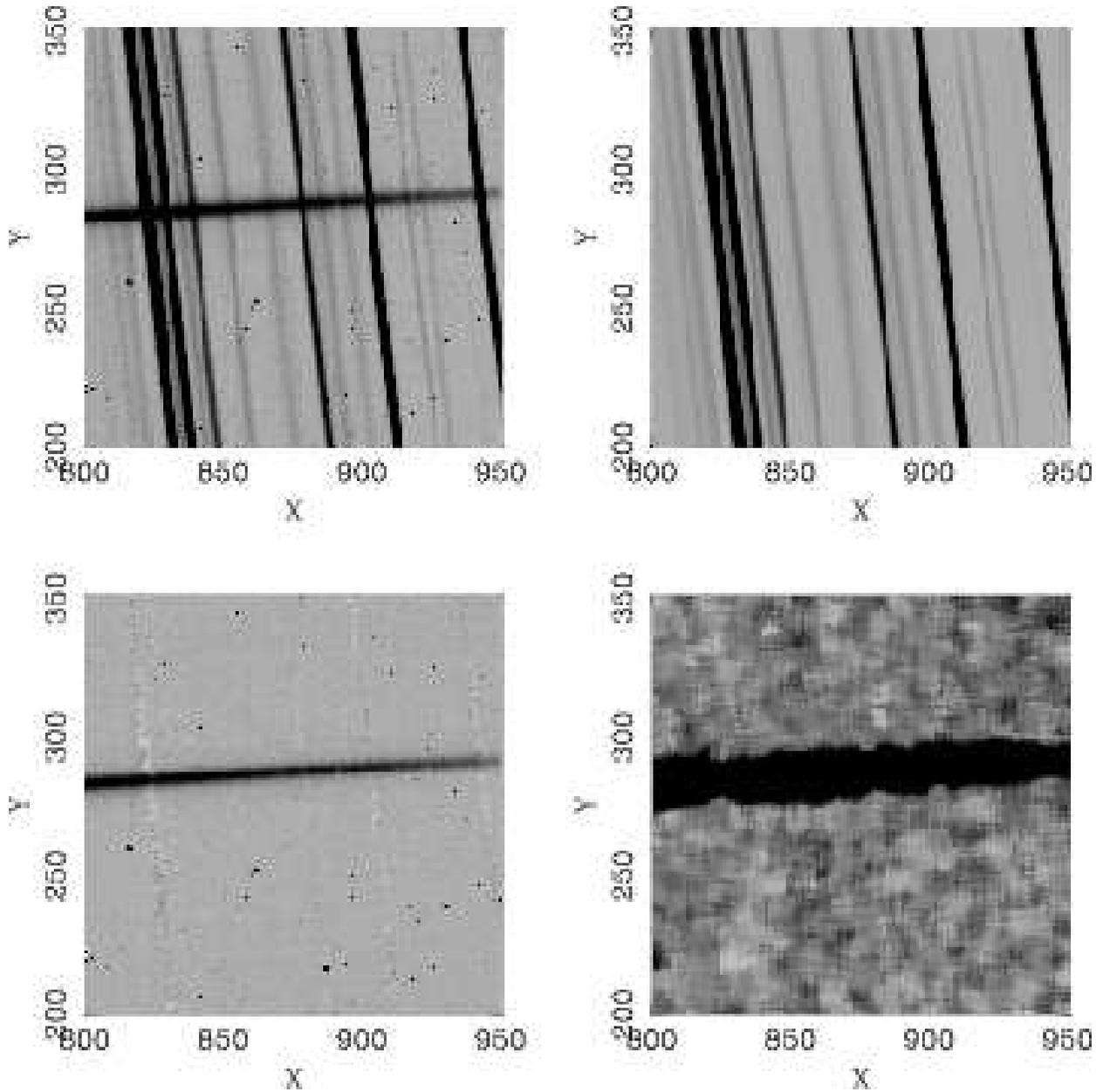}
\caption{A section of the data in Figure \ref{fig:nirspec1}. Note the
coarse sampling (first panel). The second panel shows the model background
spectrum for the same section, in which sampling in the model is identical. As
discussed in the text, it is the exact reproduction of the sampling that
allows one to remove the sharp features in the background with great
accuracy. Extraction of the object spectrum can be performed with or
without rebinning the data, depending on the needs of the individual
observer.
\label{fig:nirspec2}
}
\end{figure}

\begin{figure}
\plotone{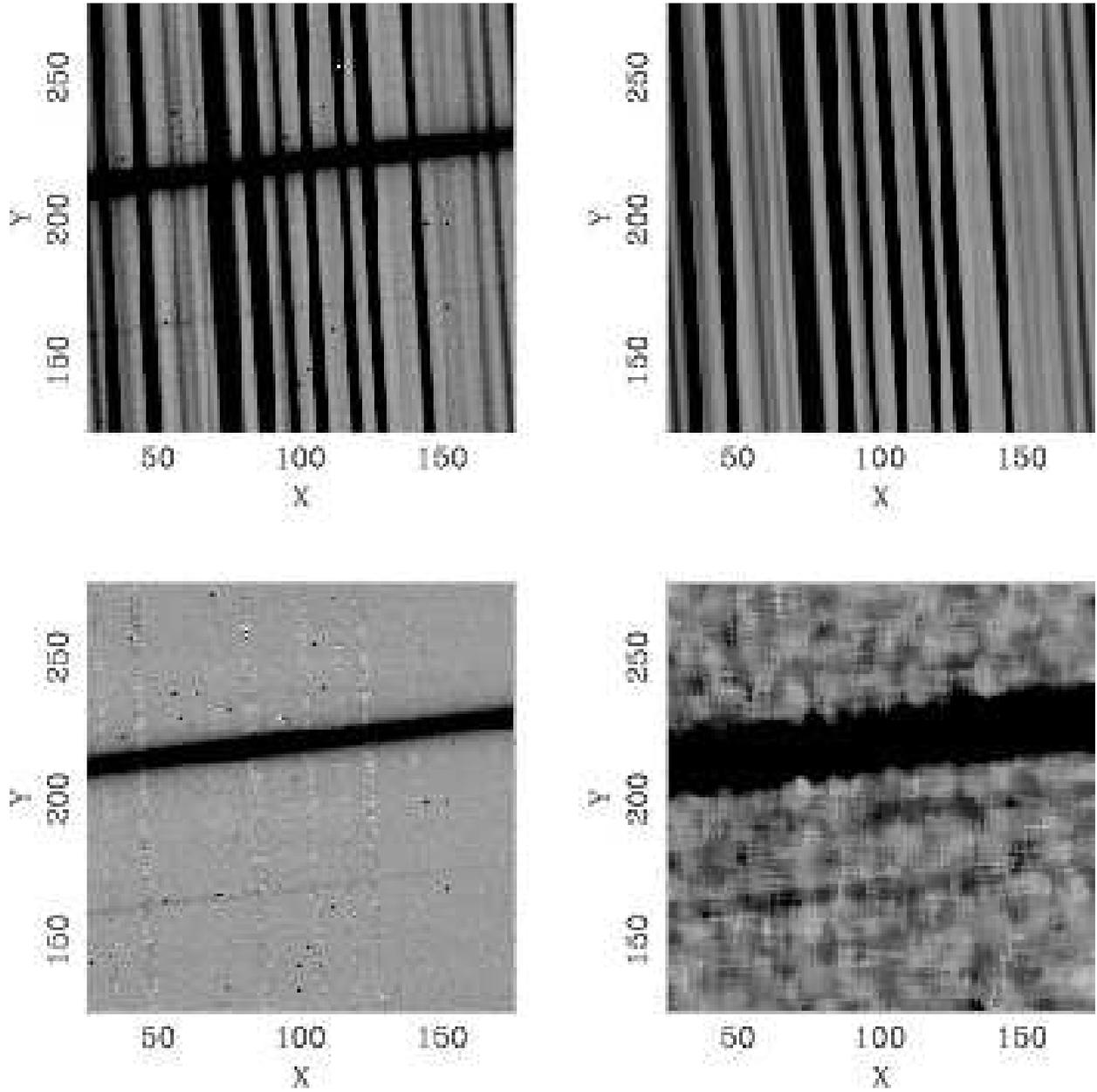}
\caption{Same as in Figure \ref{fig:nirspec2}, but for a different
section of the NIRSPEC data.
\label{fig:nirspec3}
}
\end{figure}

Figure \ref{fig:nirspec1} shows a two-dimensional $H$-band spectrum
obtained with NIRSPEC \citep{nirspec}. The dispersion is approximately
2.8\AA/pixel. The distortions in the data are large, and the sampling is
poor. Small sections of these data are also shown in Figures
\ref{fig:nirspec2} and \ref{fig:nirspec3}. In this example, the knots in
the dispersion direction were located at half-pixel intervals. This choice
was motivated by the strong distortions and length of the slit. Together
these allow for higher resolution in the reconstructed background spectrum.

\begin{figure}
\epsscale{1.0}
\plotone{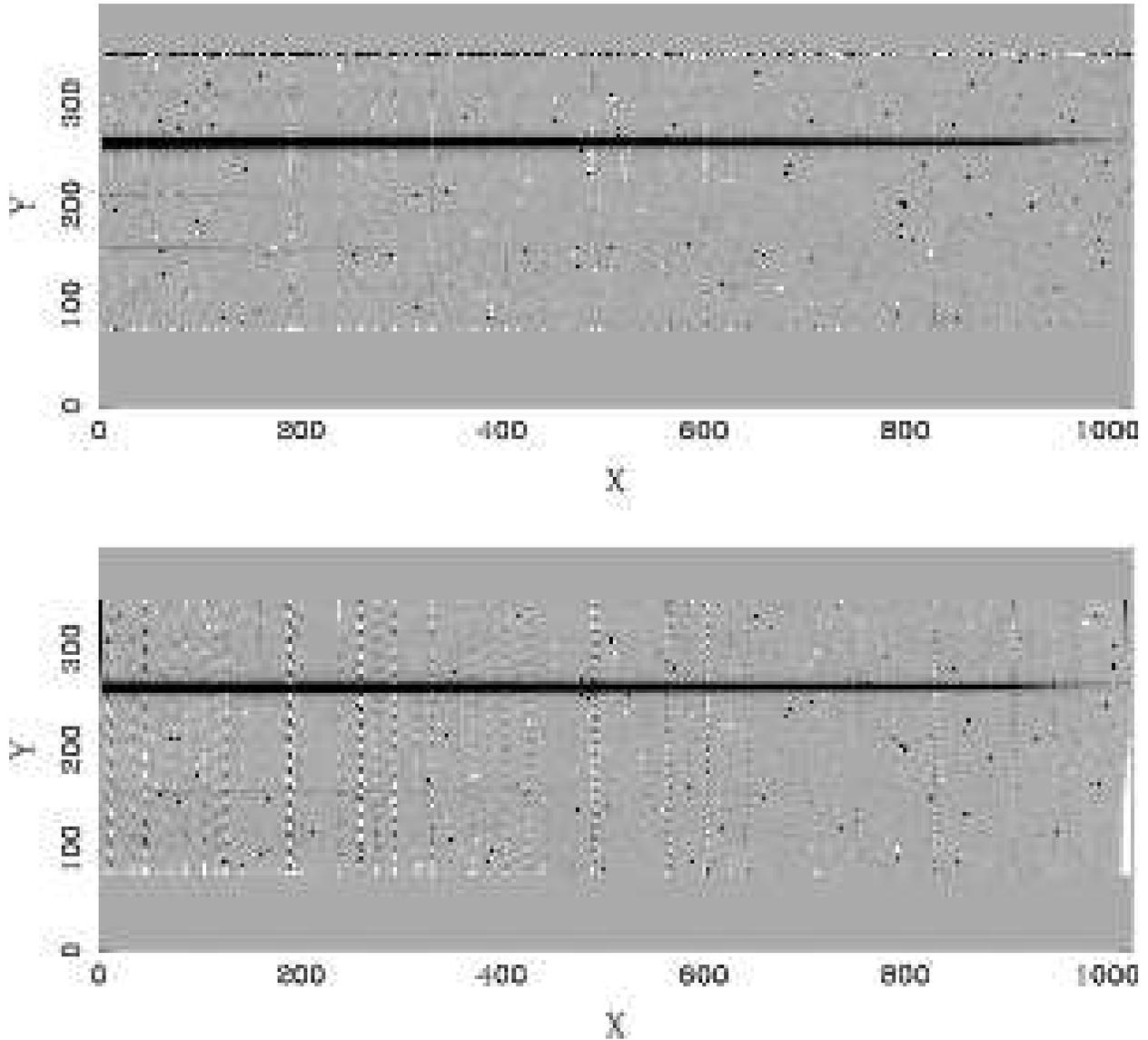}
\epsscale{1.0}
\caption{(top) Rectified two-dimensional sky-subtracted spectrum
from Figure \ref{fig:nirspec1}, where the rectification was performed {\it
after\/} the task of sky-subtraction. (bottom) The same data frame but
where the sky subtraction was performed {\it after\/} rectification of the
data. By rebinning the under-sampled night sky lines, one is left with
periodic artifacts in the data. In the top panel the night sky background
was subtracted from the original unrebinned data prior to rectification. As
a result, the noise in the sky is rebinned but not the sharp features
themselves.
\label{fig:nirspec4}
}
\end{figure}

\begin{figure}
\epsscale{1.0}
\plotone{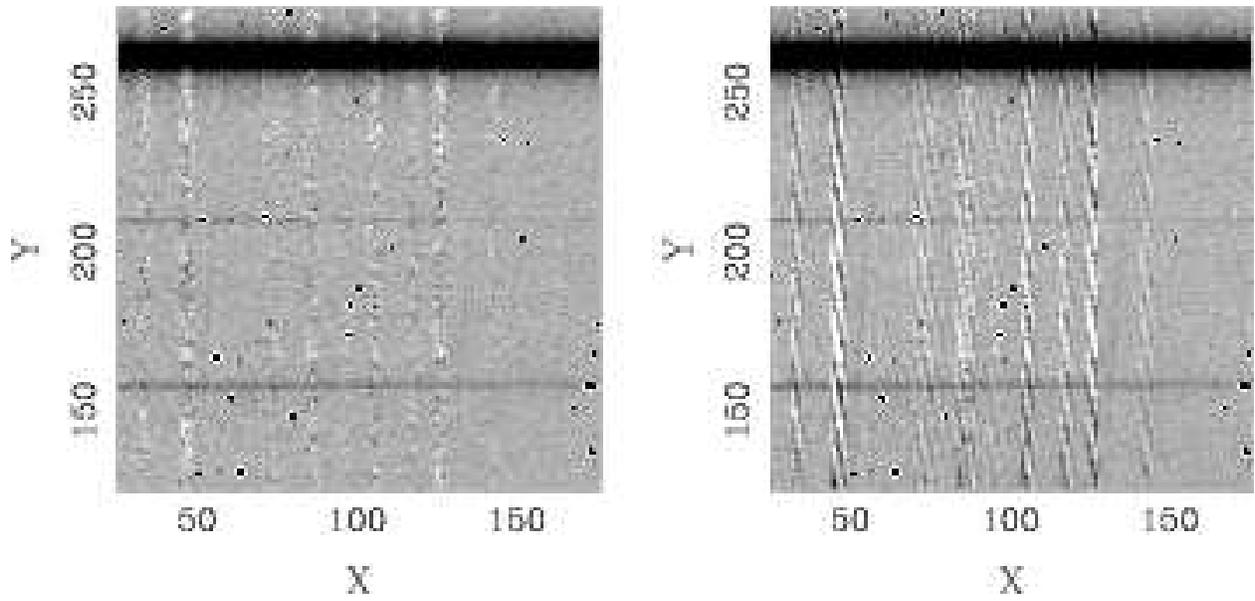}
\epsscale{1.0}
\caption{(left) A subsection of the rectified data, where the task of
sky-subtraction was performed first; (right) a subsection of the rectified
data, where the task of sky-subtraction was performed after rebinning the
data. Undesired artifacts are clearly visible in the right-hand panel
at the locations of the under-sampled night sky emission lines.
\label{fig:nirspec5}
}
\end{figure}

Even in coarsely sampled near-IR spectra, the method provides a
two-dimensional model of the background with identical sampling as in the
original data. Thus, the sky background subtracts cleanly, with no
systematic residuals at the locations of the night sky emission lines. When
the sampling is this coarse, the observer should avoid interpolating sharp
features, and observers now can choose whether to rebin the sharp features
in the raw data or to rebin the sky-subtracted frame. Figures
\ref{fig:nirspec4} and \ref{fig:nirspec5} employ these NIRSPEC data to
directly compare this new methodology for sky-subtraction with traditional
procedures. In the figures, we show the NIRSPEC data in two states: (1)
sky-subtracted first and rectified second; and (2) rectified first and
sky-subtracted second. In the latter form, strong, periodic artifacts are
plainly visible at the locations of the night sky emission lines, where the
under-sampled edges were rebinned. In the former, where the task of
sky-subtraction was performed first, no sharp gradients remain in the data
to introduce artifacts into the rectified frame. While this example shows a
comparison of rectified two-dimensional sky-subtracted data, some readers
may choose not to rebin their two-dimensional spectra at all, and opt for
extracting one-dimensional spectra directly from the unrectified
background-subtracted frames (using knowledge of the two-dimensional
wavelength solution).

\subsection{MIKE}

\begin{figure}
\epsscale{1.0}
\plotone{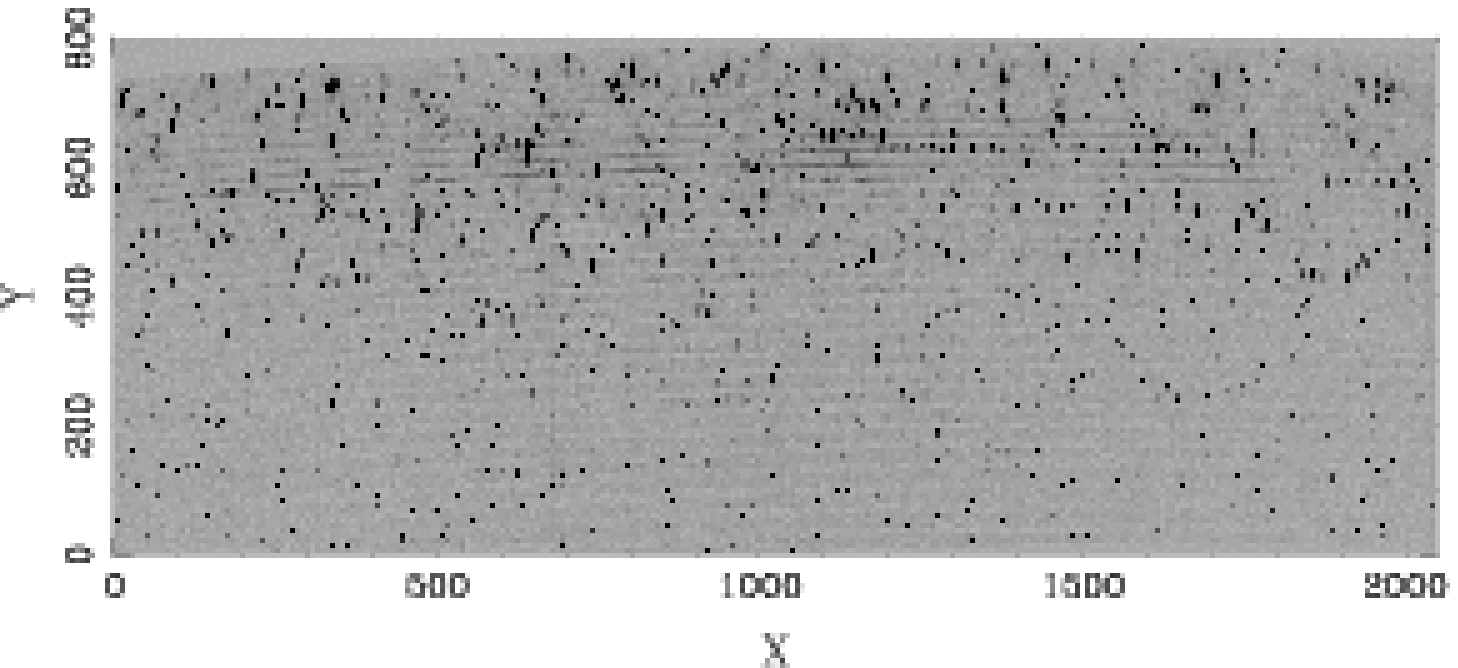}
\plotone{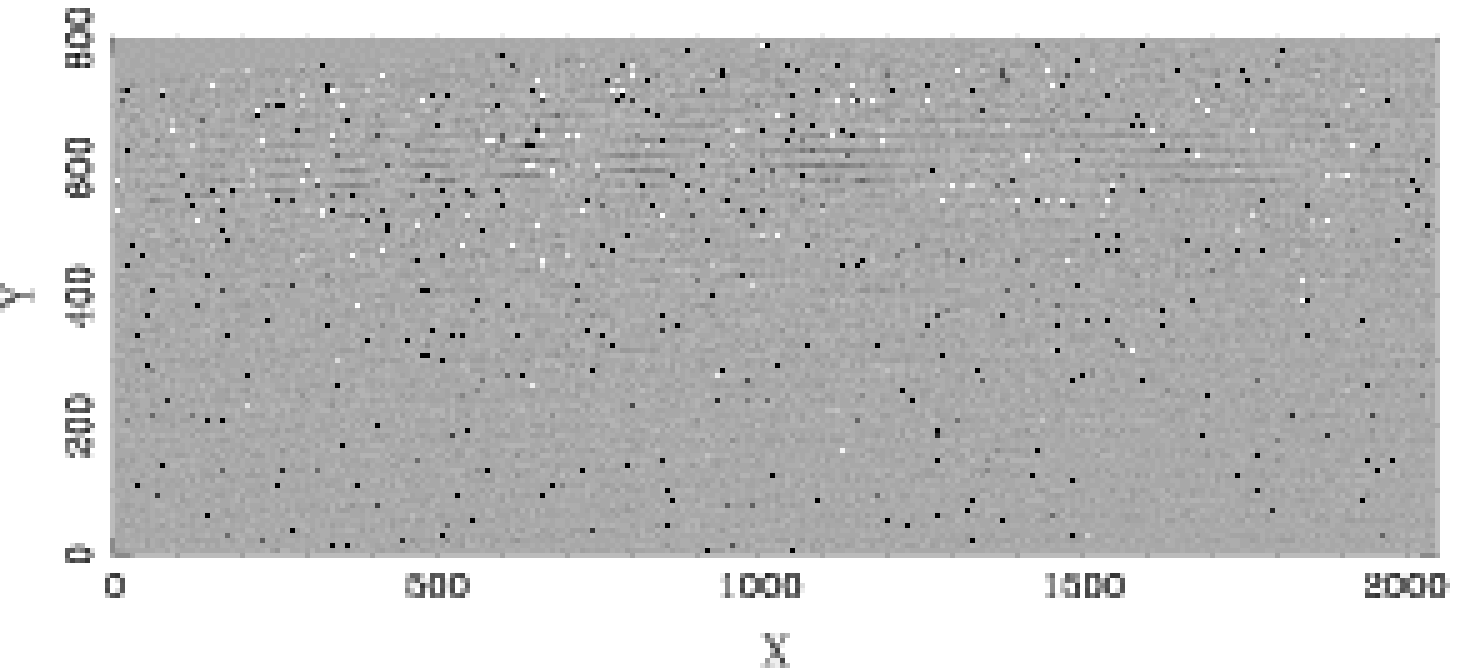}
\epsscale{1.0}
\caption{(a) One hour exposure of a QSO at $z=5.8$ using the red side
of the MIKE echelle spectrograph on the Clay telescope at Magellan.
This exposure covers order \#62 (bottom, central wavelength $\sim 5540$\AA)
through order \#33 (top, central wavelength $\sim 10300$\AA). The data were
binned $2\times 2$, effectively increasing the fraction of the image
contaminated by cosmic rays by a factor of four. The binned data also
have a dispersion of $\sim 0.1$\AA/pixel. Because the data were binned, the
night sky emission lines are heavily under-sampled and traditional
rectification and sky subtraction techniques introduce artifacts into
the data (see Figure \ref{fig:mike2}). (b) Same as (a) but with the
background subtracted from it.
\label{fig:mikered}
}
\end{figure}

\begin{figure}
\epsscale{1.0}
\plotone{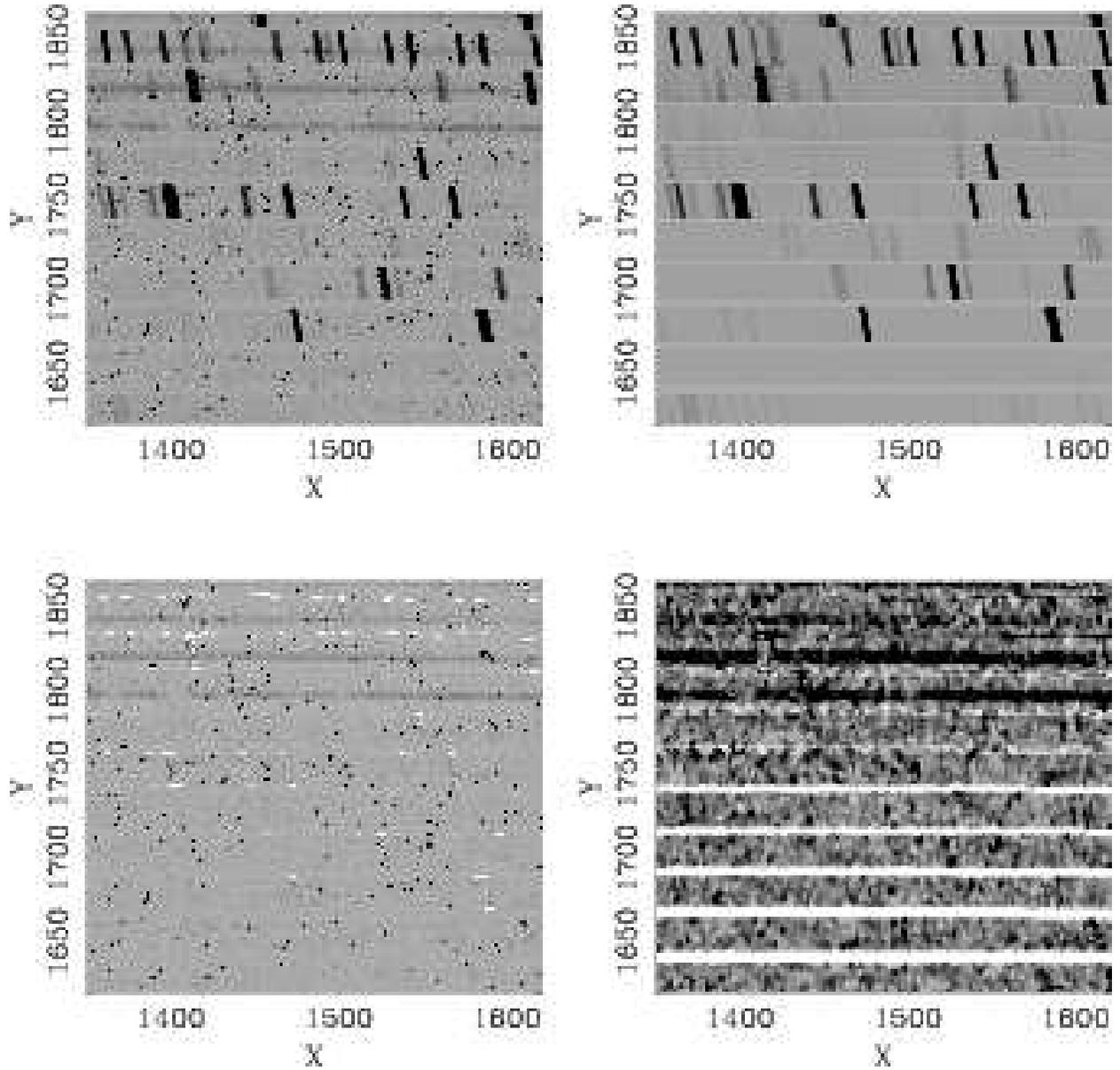}
\epsscale{1.0}
\caption{A subsection of the data shown in Figure \ref{fig:mikered}. This
section of the data has little order curvature, but the line curvature is
clearly visible. With accurate maps of the distortions and line curvature,
the bivariate B-spline accurately recovers the night sky spectrum at
wavelength intervals smaller than a pixel, leaving only the object
spectrum, cosmic-rays, and Poisson noise.
\label{fig:mike1}
}
\end{figure}

\begin{figure}
\epsscale{1.0}
\plotone{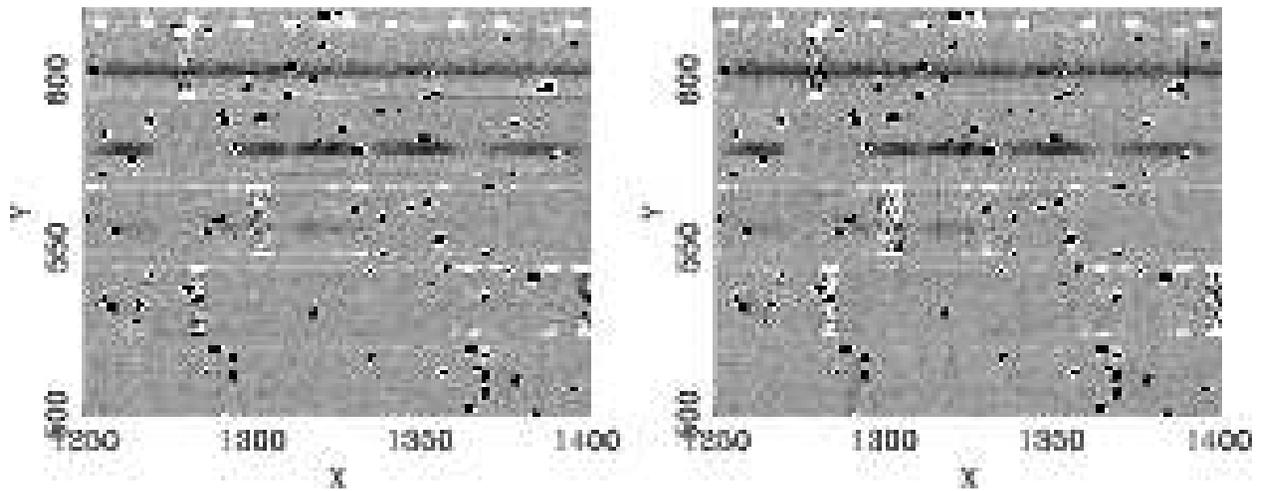}
\epsscale{1.0}
\caption{
(left) A section of the rectified two-dimensional sky-subtracted spectrum
from Figure \ref{fig:mikered}, where the rectification was performed {\it
after\/} the task of sky-subtraction. (right) The same data frame but
where the rectification was performed {\it before\/} sky subtraction. By
rebinning the under-sampled night sky lines, one is left with periodic
artifacts in the data. By rectifying the sky-subtracted frame, one rebins
the noise in the sky but not the sharp features themselves. Of course
if one is only interested in one-dimensional spectra, then the data need
not ever be rebinned.
\label{fig:mike2}
}
\end{figure}

Figures \ref{fig:mikered} -- \ref{fig:mike2} show data from a one hour
integration with the red side of the MIKE echelle spectrograph on the Clay
6.5m telescope at Magellan. Because the optical path in MIKE \citep{mike}
involves using one prism as the primary disperser and the cross-disperser,
the orders have large curvature and the spectra themselves show large line
curvature. This exposure covers order \#62 (bottom, central wavelength
$\sim 5540$\AA) through order \#33 (top, central wavelength $\sim
10300$\AA). The data had been binned $2\times 2$ while reading the CCD to
avoid being read-noise dominated, even with the one hour integration. The
binning increases the fraction of the image contaminated by cosmic rays by
a factor of four, and reduces the spatial extent of the slit to about 20
pixels. Many users also use $1\times 3$ and even $2\times 4$ binning with
MIKE, and as a result, most data obtained with MIKE will be severely
under-sampled, and the binning leads to higher contamination rates with
cosmic rays.

In Figure \ref{fig:mike2} we show a direct comparison of a portion of the
rectified sky-subtracted MIKE data (similar to the comparison for NIRSPEC
data in Figure \ref{fig:nirspec4}). While at such high resolution the night
sky lines do not affect much of the data, one may still wish to recover
accurate estimates of the object flux at the locations of the sky lines. In
the right-hand panel of Figure \ref{fig:mike2} there are clear artifacts at
the locations of the sky-lines and such artifacts render the science data
useless at those wavelengths. In the left-hand panel, the object spectrum
is not contaminated by the artifacts introduced by the rebinning of night
sky lines.

\bigskip

The examples shown in this section illustrate a range of results from the
modeling procedure outlined in this paper. In all cases, when the
distortions and line curvature are accurately known, the sky background
shows very little spatial variation when fit in the rectified coordinate
system (see \S \ref{sec:proc}). The need for a high spatial order is only
required when using traditional methods, and artifacts at the edges of sky
lines are not well modeled with low-order polynomials. Because of this,
observers can also apply this method to short slitlets using one-sided sky,
restricting the B-spline to one dimension (wavelength) or perhaps only
limiting the spatial order to $k_y=1$. In such cases, the only factor
limiting the accuracy of the sky subtraction would be flat-fielding and
other techniques may prove more valuable \citep[e.g.,][]{nod}.

\section{Summary}

This paper describes a new method for modeling the background in
two-dimensional slit spectroscopy.\footnote{An implementation of the method
is available as part of a suite of Python reduction scripts written by the
first author at {\tt http://www.ociw.edu/$\sim$kelson\/}. Though no
documentation currently exists, readers may download at their discretion.}
This method makes full use of the data in its ``raw'' state --- before the
data have been rectified (rebinned). The distortions inherent in the data
provide the means by which the background spectrum can be accurately
modeled at repeated sub-pixel locations. Because of the improved sampling
of the background spectrum in the original data compared to a rectified
image, one can also robustly reject pixels contaminated with cosmic rays
and bright objects. Because of these advantages over traditional
background-subtraction techniques, the task of subtracting the
two-dimensional background spectrum should now be performed as one of the
first steps in any pipeline of spectroscopic reduction, even before cosmic
rays and other bad pixels have been ``cleaned''.

The subtraction of the background can now be performed on raw spectra,
riddled with cosmic rays. As a result, the procedure can be adopted as
one of the first steps in a pipeline for spectral reduction, and not as
one of the last. The task is computationally inexpensive and can allow
observers to quickly analyze incoming data at the telescope, assuming
the distortion maps have been characterized from calibration data during
the previous afternoon, or are known {\it a priori\/} from an optical
model of the instrument. After the background has been subtracted the
cosmic rays can be flagged trivially, without the need for complicated
cosmic ray identification procedures.

Because the algorithm is utilizing the full sampling of the background
spectrum produced by the distortions in the raw data, the method is
making optimal use of the available data. While this procedure has been
presented in the context of taking full advantage of the optical
distortions and spectral line curvature to recover better sampling of
areas with strong gradients in the data, the method works just as well
on data for which the line curvature is negligible. This recovery of the
two-dimensional background spectrum leaves no systematic residuals at
the locations of strong gradients simply because the model is never
interpolated at locations where the data do not exist. The model is
always sampled where the original data exist. As a result, the sharp
gradients, such as those at the edges of bright sky emission lines, are
never rebinned and ringing does not occur.

For spectrograph/detector combinations that over-sample the data (e.g.,
DEIMOS and the IMACS long camera), traditional methods may be deemed
satisfactory by many observers. However, in near-IR or high-resolution
echelle spectroscopy, the data may be binned and/or under-sampled in order
to reduce the importance of read noise. Regardless of the source of one's
spectroscopic data, observers now can choose whether to rebin their data
{\it with\/} or {\it without\/} the sharp night-sky features present.
Observers may now choose the latter, as the rebinning of coarsely sampled
features produces unwanted artifacts that prevent the accurate modeling of
the background, and prevent the accurate recovery of an object's flux at
every observed wavelength.

Nevertheless, for observers who wish to proceed from the telescope to
extracted spectra in the fewest number of steps possible, without the
additional correlated noise introduced by rebinning ones data, and without
any artifacts caused by rebinning any strong gradients, computational
machinery has now caught up to your demands.

\acknowledgements

The author would like to thank the staff at the Carnegie Observatories for
their support, and in particular P. Martini and M. Rauch for the use of
their NIRSPEC and MIKE data. Furthermore, the anonymous referee is
acknowledged for ensuring that this work is received by a broader audience.



\clearpage

\begin{thebibliography}{}

\bibitem[Bernstein et al.(2003)]{mike} Bernstein, R. et al. 2003,
\procspie, in press

\bibitem[de Boor(1978)]{splinebook} de Boor, C., Applied 
Mathematical Sciences, New York: Springer, 1978

\bibitem[Dierckx(1993)]{dierckx} Dierckx, P., Curve and Surface
Fitting with Splines, New York: Oxford University Press, 1993

\bibitem[Glazebrook \& Bland-Hawthorn(2001)]{nod} 
Glazebrook, K.~\& Bland-Hawthorn, J.\ 2001, \pasp, 113, 197 

\bibitem[Kelson \etal (2000)]{expector}Kelson, D.D., Illingworth, G.D.,
van Dokkum, P.G., \& Franx, M. 2000, \apj, 531, 159

\bibitem[Kurtz \& Mink(2000)]{svdfit} Kurtz, M.~J.~\& Mink, 
D.~J.\ 2000, \apjl, 533, L183 

\bibitem[McLean et al.(1998)]{nirspec} McLean, I.~S.~et al.\ 1998,
\procspie, 3354, 566 

\bibitem[Oke \etal (1995)]{okelris}Oke, J.B., \etal\ 1995, \pasp, 107, 375

\bibitem[Viton \& Milliard(2003)]{viton} Viton, M.~\& 
Milliard, B.\ 2003, \pasp, 115, 243 



\end{thebibliography}
\end{document}